# Correlating chemical composition and optical properties of photochromic rare-earth oxy-hydrides using ion beam analysis

S. M. Aðalsteinsson[1,*], M. V. Moro[1], D. Moldarev[1,2], S. Droulias[3], M. Wolff[1,2] and D. Primetzhofer[1,3]

[1]Department of Physics and Astronomy, Uppsala University, Box 516, 751 20 Uppsala, Sweden
[2]Department of Material Science, Moscow Engineering Physics Institute, 115409s Moscow, Russia
[3]Tandem Laboratory, Uppsala University, Box 529, 75120 Uppsala, Sweden

**Abstract**

We relate the photochromic response of rare-earth oxy-hydride thin films (YHO, NdHO, GdHO and DyHO) synthesized by reactive magnetron sputtering to chemical composition. Depth profiles of the sample composition are extracted by a multi-method ion beam analysis approach. The total areal density of the thin films is deduced from Rutherford Backscattering Spectrometry while coincidence Time-of-Flight/Energy Elastic Recoil Detection Analysis provides depth-profiles of the film constituents. High-resolution depth profiles of the concentration of light species, i.e. hydrogen and oxygen, are additionally extracted from Nuclear Reaction Analysis and Elastic Backscattering Spectrometry, respectively. The photochromic response of the films is measured by optical transmission spectroscopy before and after illumination. We report photochromic properties for YHO, NdHO, GdHO and DyHO for chemical compositions described by the formula $REH_{2-\delta}O_\delta$ in the range of $0.45 < \delta < 1.5$.

*Corresponding author:
Sigurbjornmar.Aealsteinsson.2150@student.uu.se

## 1. Introduction

Recently, it has been observed that thin films of yttrium oxy-hydride (YHO) can exhibit a reversible photochromic response at ambient conditions, transforming from a transparent to a darkened state under illumination by visible light [1]. This process is accompanied by an increase of conductivity [2]. These distinct properties make the material attractive for technological applications, for instance, next generation smart-windows, optoelectronics and hydrogen-storage devices [3-5].

Transparent and photochromic YHO thin films can be produced using reactive magnetron sputtering in an Ar:$H_2$ atmosphere, followed by oxidation when the samples are exposed to air [6]. It was shown that the photochromic response can be tuned by the total deposition pressure [7]. C. You and co-authors showed that the optical properties of YHO thin films are strongly dependent on the level of oxygen incorporated in the films. At a certain threshold oxygen concentration, the thin film transforms from an opaque, metallic state into a transparent and photochromic state. Further increase in oxygen concentration results in decreased photochromic response, while the band gap was found to increase [8]. Even higher oxygen concentrations result in the formation of an optically inactive transparent oxide. In a more detailed study of composition by D. Moldarev et al. [9] it is was revealed that oxygen replaces hydrogen during the oxidations process and follows the chemical formula $YH_{2-\delta}O_\delta$. For samples with $0.45 < \delta < 1.5$ photochromic properties are reported. *In-situ* analysis of the samples under illumination showed that photochromism can be triggered in high-vacuum conditions as well and is not related to significant changes of the integral composition of the film ( than ≈1at. %) [10]. This finding suggests that the photochromism might be related to either microstructural changes or electronic rearrangements without substantial modification of the composition. Evidence for light-induced lattice contraction was earlier reported for photochromic YHO samples [11], accompanied by changes of vibrational modes [12].

More recently it was shown that a reversible photochromic effect is observed in other rare-earth-based oxy-hydrides (REHO, RE = Gd, Dy and Er) [13] as well and the chemical composition of some of the photochromic films (YHO, ScHO and GdHO) was related to changes in bandgap [14].

In this work, we synthesized a series of different REHO (RE = Y, Nd, Gd and Dy) thin films by reactive magnetron sputtering, and determine their photochromic response using optical spectroscopy. Subsequently, their respective chemical composition was deduced using a multimethod IBA approach to correlate the photochromic response to chemical composition. We report the observation of composition-dependent photochromism for all investigated systems, including NdHO, in the composition range of $0.4 < \delta < 2$. For NdHO thin films, to the best of our knowledge, photochromism has not been reported before, while synthesis processes for NdHO powder were demonstrated [15].

## 2. Methodology

### 2.1 Sample preparation

The REHO thin film samples were produced by reactive argon-magnetron sputtering using a compact Balzers Union sputtering system at Uppsala University. For reactive sputtering hydrogen gas is introduced into the sputtering at controlled partial pressure. Commercially available Y, Dy, Gd and Nd sputtering targets were used, with nominal bulk purity 99.99%, 99.9%, 99.9%, and 99.9%, respectively. The thin films were deposited onto soda-lime glass substrates (microscope slides, 10x10 mm$^2$ and 1 mm thick). The base pressure in the chamber before deposition was lower than 7x10$^{-3}$ Pa. Thin films were grown at room temperature in an Ar:H$_2$ environment (H$_2$/Ar ratio variated from 0.01 to 0.09), and the pressure during deposition in the range of 0.6 - 0.9 Pa following the procedure described elsewhere [16]. The target-substrate distance was varied from 4 cm to 8.5 cm and the plasma sputtering current was 120 mA. Subsequently to deposition, samples were exposed to air for oxidation. Initially the sputtered films are in a form of an opaque yttrium hydride (YH$_x$), and upon exposure to air transform to YHO with a yellowish transparent appearance [17].

### 2.2 Optical characterization

Optical measurements of the REHO samples were performed using a Perkin Elmer Lambda 35 UV/Vis spectrophotometer, equipped with tungsten and deuterium light sources. The transmission measurements were calibrated with respect to 100 % transmission in air. The spectrophotometer was operated in scanning-mode in the wavelength range [300-1000] nm at a speed of 240 nm/min (slit size 2 mm). Initially, the optical transmission is measured while the samples were fully bleached (i.e., initial stage before illumination), followed by a measurement after 2 hours of illumination using a LED lamp (wavelength 400 nm and intensity ≈ 10 mW/cm$^2$). All samples were found to relax to their original state once illumination was stopped. Figure 1 (a) illustrates the photo darkening effect for a DyHO sample. In this work, we define photochromic response as the ratio of the averaged optical transmission contrast of the spectra before and after illumination (integrated within the wavelength range [500 – 900] nm) to the transmission before illumination. A typical optical transmission spectrum, before (dashed line) and after illumination (continuous line), is shown in Fig. 1(b), for a DyHO sample.

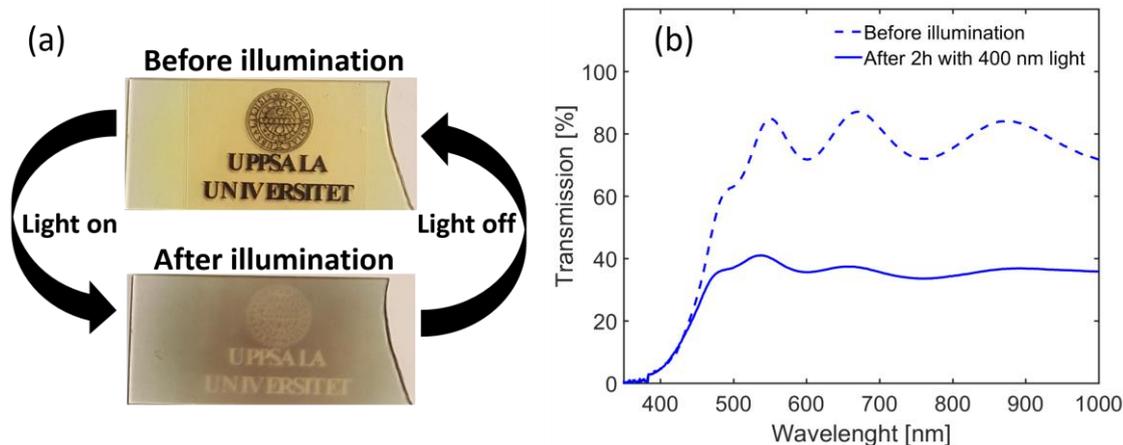

**Figure 1:** (a) Photochromic response of dysprosium oxyhydride before and after 2h of illumination. (b) Optical transmittance spectra showing the corresponding reduction in transmittance of the same sample shown in (a).

### 2.3 X-ray diffraction

X-ray diffraction was utilized to investigate the structure of the photochromic NdHO thin film (for other photochromic films, see references [12,14]), using a Siemens D5000 diffractometer at Uppsala University with a Cu-Kα X-ray source (λ = 1.540 Å) in 2θ ranging from 15 to 80 degrees, in Bragg-Brentano geometry at ambient conditions.

### 2.4 Ion Beam Analysis

The chemical composition of the REHO thin films was investigated by four IBA methods: Rutherford Backscattering Spectrometry (RBS), Time-of-Flight Elastic Recoil Detection Analysis (ToF-ERDA), Nuclear Reaction Analysis (NRA), and Elastic Backscattering Spectroscopy (EBS) [18]. The IBA measurements were performed at the Tandem Laboratory at Uppsala University using the 5-MV NEC-5SDH-2 tandem accelerator [19].

RBS measurements were done with a 2 MeV He[+] beam. The backscattered particles were detected using a planar ion-implanted passivate detector (PIPS) fixed at = 170° scattering angle, with a resolution FWHM of ≈ 13 keV, covering a solid angle of (2.16 ± 0.11) msr. From RBS the elemental composition of the metallic elements was extracted, as this method is generally more sensitive to heavier elements [18].

EBS measurements were carried out using the elastic $^{16}O(\alpha, \alpha_0)^{16}O$ resonance at 3.037 MeV He[+] to investigate the concentration of oxygen with high accuracy [20]. EBS spectra were measured in two energy regimes probing oxygen close to the surface and in the bulk of the films.

For the ToF-E ERDA measurements, the ToF-E telescope tube is fixed at 45° degrees and features an ionization gas detector chamber (GIC) for energy discrimination (details of the setup can be found elsewhere [21]). The coincidence ToF-E ERDA measurements were performed using 36 MeV I$^{8+}$ as primary beam directed onto the sample under an incident angle of 67.5° with respect to the sample normal. The constituents of the sample are recoiled towards the coincidence time-of-flight and energy detection system allowing for a mass-resolved composition depth profiling of the sample [22].

NRA measurements were performed using a lead-shielded large solid angle gamma detector. High resolution hydrogen depth profiling was performed using the $^{1}$H($^{15}$N,αγ)$^{12}$C nuclear reaction at 6.385 MeV [23]. Hydrogen depth profiling was achieved by tuning the incident ion energy away from resonance. The characteristic 4.44 MeV γ-emission from the $^{1}$H($^{15}$N,αγ)$^{12}$C nuclear reaction is normalized via a reference of hydrogen implanted silicon with stable and well-known concentration and depth (details on the evaluation procedure can be found elsewhere [24]).

**3. Results and discussion**

### 3.1. Iterative composition characterization

All samples were characterized by RBS and ToF-ERDA directly after growth. Fig. 2 (a) shows RBS spectra for a photochromic DyHO sample (black dots for experimental data and red solid line for a fit using SIMNRA [25]). The Dy signal is well separated, whereas identification and quantification of lighter constituents in the film is difficult due to scattering from the substrate. To quantify the lighter elements, panel (b) depicts a coincidence ToF-E ERDA spectrum for the same sample (the inset shows the depth-profile derived using Potku [26]). We find C contamination (≈ 5 at.%) throughout the film, as well as a Si signal from the silica glass substrate. The presence of small quantities of contaminants such as C and N is observed in all materials investigated. Small amounts of F (≈ 2 - 4 at.%) were only found in the YHO thin films, probably from contaminations of the sputtering target.

ToF-E ERDA depth-profiles require accurate knowledge of stopping powers and efficiency corrections of the ToF-E system and thus generally allow only an accuracy of about 10–20% in relative concentrations. An additional source of uncertainty can be depletion of hydrogen during the analysis due to heavy-ion bombardment [27]; see for instance Fig. 2 in Ref. [28] for YOH. However, the relative concentrations extracted for similar samples at similar doses are more accurate (see discussions in Ref. [29]). For this reason and to provide a reference point for each batch of thin films, one sample from each batch was studied in addition by NRA and EBS.

In fig. 2 (c), the NRA spectrum of the same DyHO sample is shown. The energy region used for integration of the gamma emission is marked by dashed lines. The inset depicts the hydrogen concentration extracted from the data as a function of depth. The average hydrogen concentration is around ≈ 14 at. %, somewhat constant throughout the film, consistent with previous findings [9]. Possible ion-induced hydrogen-desorption [30] was studied by performing multi-scans in list-mode data acquisition, and no significant losses of hydrogen (≈1-3 at.%) were identified. The H depth-profile obtained from NRA was used to fine-tune the H results from the ToF-E ERDA analysis analysis (i.e., inset panel (b)). The difference between both H profiles was found to be on average ≤ 20 % (in agreement with our expectations - see discussions in [29]).

Finally, in panel (d), a typical EBS spectrum for 3.06 MeV He$^+$ is presented (black dots), together with the fit using SIMNRA [25] (red solid line). For the EBS fits, the SigmaCalc cross sections [31] were used as input values. The stopping power data used as input (for all ion beam analysis) was retrieved from the last version of the SRIM code [32]. The amount of O fitted to all EBS spectra is in good agreement with values deduced from ToF-E ERDA within ≈ 5 % in average.

We underline that the final RBS and EBS fits, as well as the final depth-profile from the ToF-E ERDA presented in Fig. 2, were obtained as a result of the inter-comparison amongst all IBA techniques following an iterative self-consistent analysis [33-35]. More details on the data-analysis procedure, using similar experimental setups, is discussed elsewhere [36].

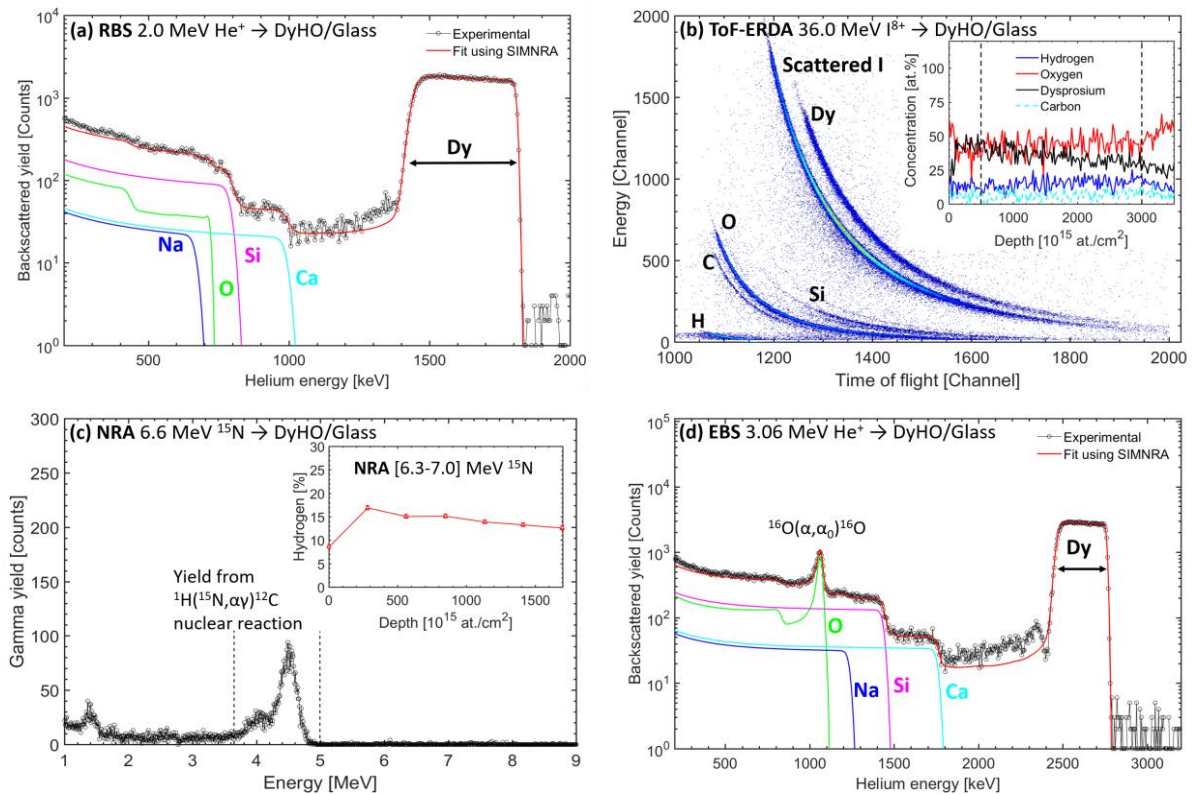

**Figure 2**: (Color Online). Data for the one DyHO thin film sample: (a) RBS spectrum (black dots) and a fit using SIMNRA (red solid line). (b) ToF-ERDA spectrum. The inset shows the atomic concentrations

of the main constituents as a function of depth. (c) NRA spectrum (black dotted-line), dashed lines indicate the region over which the gamma yield is integrated. The inset shows the deduced hydrogen depth profile. (d) EBS spectrum (black dots) recorded at beam energy of 3060 KeV, and a fit using SIMNRA code (red solid line).

### 3.2. Correlating composition and optical response

Figure 3 shows the ternary diagram of the RE (RE = Y, Nd, Gd and Dy), H and O concentrations of all the REHO samples investigated in this work. Concentrations were depth-integrated over the region indicated by the dashed lines in the inset of Fig. 2 (b). The thin oxygen-rich layer on the surface of the samples is excluded and contaminants in the film such as C, N, F are not taken into account. YHO, NdHO DyHO and GdHO are represented by blue squares, black diamonds, green circles and red triangles, respectively. The concentration of the rare-earth metal constituents is constant around 33 at.% (horizontal dashed-line), whereas O replaces H from nearly stoichiometric $REH_2$ to almost $REO_2$. These results are in very good agreement to previous work [9, 14]. In Ref. [9], the authors proposed the chemical formula $YH_{2-\delta}O_\delta$. Our findings suggest that the investigated Gd, Dy, and Nd oxy-hydrides follow the same trend.

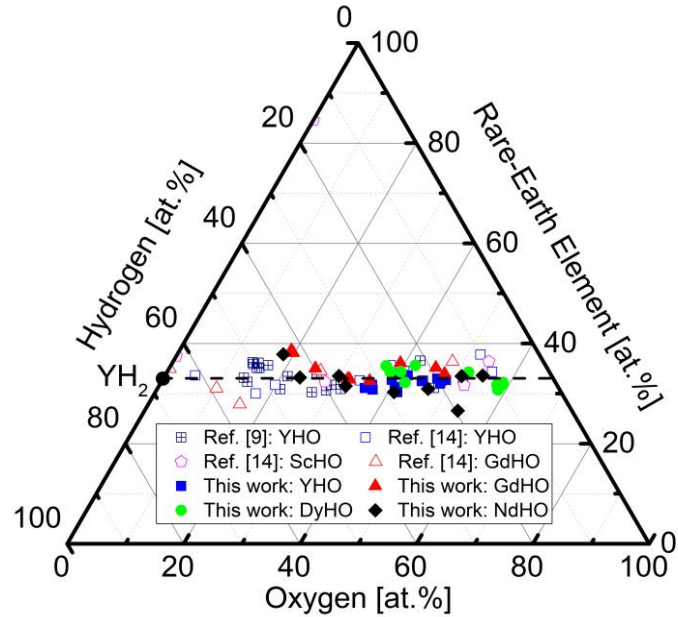

**Figure 3:** Ternary diagram of REHO (RE = Y, Gd, Dy, and Nd) for all samples investigated in this work. The dashed line represents compositions corresponding to the chemical formula $REH_{2-\delta}O_\delta$.

In Fig. 4, the photochromic response deduced from optical measurements before and after illumination for all the REHO thin films normalized by thickness is plotted as a function of chemical composition. The photochromic response was normalized by thickness (≈ 700 nm for YHO, ≈ 400 nm for DyHO, ≈ 700 nm for NdHO and ≈ 1000 nm for GdHO, in average – assuming bulk densities) to reduce possible contrast dependence. The samples are found to be transparent and photochromic in a composition range of 0.45 < δ < 1.5. A general trend of increasing photochromic response for decreasing δ is found, and it seem that a maximum of photochromic response for Gd and Nd are found around δ ≈ 1.0 and δ ≈ 0.7, respectively. However, due to lack of data for δ < 1, it is rather difficult to confirm at which δ the strongest photochromic response is observed. Below δ ≈ 0.5 NdHO is found to be opaque and dark and not photochromic. For δ > 1.5 YHO, DyHO and NdHO are transparent and not photochromic. This result is in agreement with earlier findings for YHO [9].

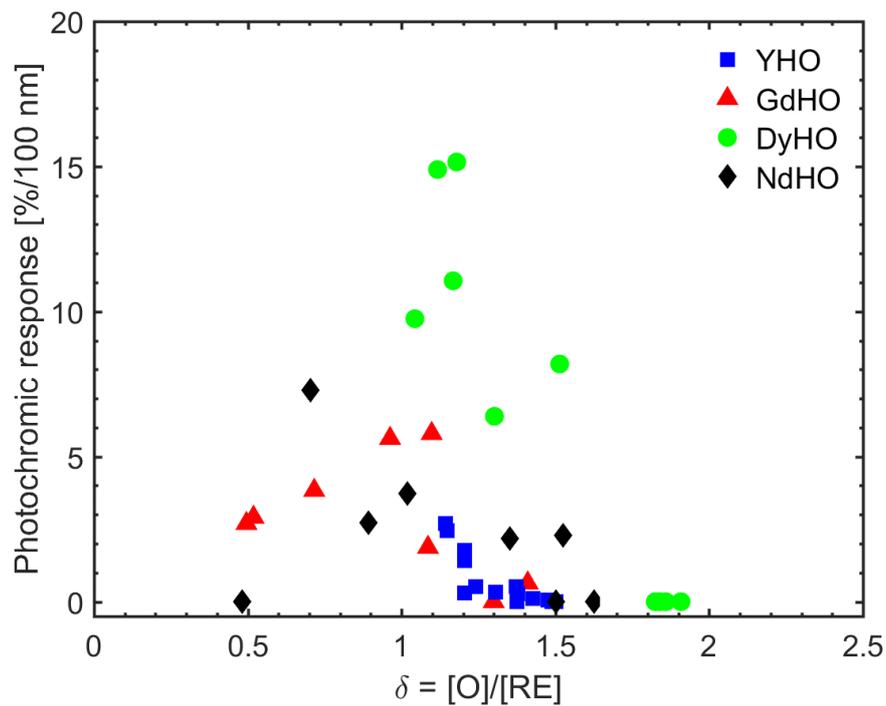

**Figure 4:** Photochromic response (averaged from 500 – 900 nm) normalized by thickness (in nm, assuming bulk densities) following 2 hours of illumination (400 nm and intensity 10 mW/cm$^2$) plotted as a function of the ratio of oxygen to rare-earth metal δ.

### 3.3 Structure of photochromic NdHO

Fig. 5 show the result of an XRD measurement of a photochromic NdHO thin film. NdHO exhibits a fcc structure (space group Fm-3m, same as REH$_2$). Earlier studies [39] using powder diffraction report that anion order/disorder structures are formed in oxyhydrides depending on the cation radius which is known to decrease with increasing atomic number (lanthanide contraction). As Nd has the largest ionic radius among the studied RE-metals, it is expected to form anion ordered (space group P4/mmm) structure in NdHO unlike YHO, DyHO and GdHO. For comparison, we indicated in Fig. 4 the diffraction peaks identified as anion ordered structure (P4/mm) and anion-disordered structure (Fm$\bar{3}$m). However, since the contribution of anionic order to the structure factor is very weak synchrotron radiation is required to draw a final conclusion.

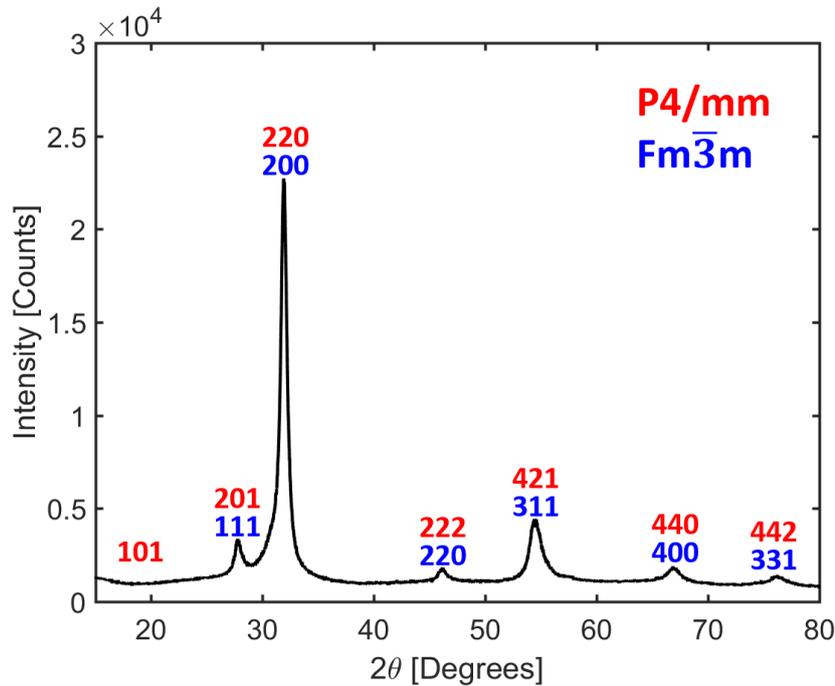

**Figure 5:** XRD powder diffraction pattern for a NdHO thin film, recorded in Bragg-Brentano geometry at room temperature (see Sec. 2.3 for further details). Other diffraction peaks are indicated in red (P4/mm) and blue (Fm3m) for comparison (see text for details).

**4. Summary and conclusions**

We have grown photochromic rare-earth oxy-hydride thin films (REHO, RE = Y, Nd, Gd and Dy) by reactive Ar magnetron sputtering followed by *ex-situ* oxidation. Ion beam-based composition analysis suggest that Y, Gd, Dy, and Nd oxy-hydrides exhibit a chemical formula close to REH$_{2-\delta}$O$_\delta$. We correlate the photochromic response to the chemical composition and show that the photochromic effect of REHO exhibit a similar correlation to their chemical composition regardless the rare-earth metal-constituent: i) films are photochromic in the compositional range 0.45 < δ < 1.5; ii) increase of oxygen concentration beyond δ = 1 leads to decrease of photochromic response. The magnitude of the photochromic response normalized by the film thickness is found similar between Y, Gd and Nd oxy-hydrides, but increases for DyHO. A systematic study on the thickness dependence of the photochromic effect is required to understand whether the photochromic effect is attributed to a surface or to a bulk effect, and thus draw a final conclusion. At this point, NdHO is already the 6[th] out of 17 rare-earth elements showing photochromic properties. Nevertheless, photochromism was reported only in elements with highest oxidation state +3, therefore future studies of elements with different oxidation state (e.g. Ce, Pr, Tb with +4) can help to gain further insights of the nature of photochromic effect.

**Acknowledgments**

The authors would like to thank Ayan Samanta at the Department of Chemistry at Ångström Laboratory for granting access to their PerkinElmer Lambda optical spectrometer and Mikael Ottosson at the Department of Chemistry at Ångström Laboratory for granting access to the XRD laboratory. D.M. acknowledges Visby Programme Scholarships for PhD studies. Support of the operation of the tandem accelerator laboratory by VR-RFI (contracts #821-2012-5144 & #2017-00646_9) and the Swedish Foundation for Strategic Research (SSF, contract RIF14-0053) is acknowledged.